 \documentclass{article}
\usepackage{graphicx}
\usepackage{times}
\usepackage{xcolor}
\usepackage{soul}
\usepackage[small]{caption}
\usepackage{graphicx}
\usepackage{subfig}
\usepackage{amssymb}
\usepackage{mathtools}
\usepackage{amsthm}
\usepackage{algorithm}
\usepackage{algorithmicx}
\usepackage{algpseudocode}
\usepackage{url}
\usepackage{threeparttable}
\usepackage[utf8]{inputenc}
\begin{document}

\begin{center}
\Large{Algorithmic Collusion in Cournot Duopoly Market: Evidence from Experimental Economics}
\end{center}

\begin{center}
\normalsize{~\\
Nan Zhou$^{1,2}$, Li Zhang$^2$, Shijian Li$^{2*}$, Zhijian Wang$^{1*}$ \\
~\\
       $^1$Experimental Social Science Laboratory, \\
       $^2$College of Computer Science and Technology,\\
       Zhejiang University, Hangzhou, 310058, China \\
       ~\\
}
\end{center}

~~\\
~~\\
~~\\

\begin{abstract}
%
Algorithmic collusion is an emerging concept.
  Whether algorithmic collusion is a creditable threat remains as an argument.
  In this paper, we propose an algorithm
  which can extort its human rival to collude in a Cournot duopoly competing market.
  In experiments, we show that, the algorithm can successfully extort its human rival
  and gets higher profit in long-run,
  meanwhile the human rival will fully collude with the algorithm.
  As a result, the social welfare declines rapidly and stably.
  Both in theory and in experiment, our work confirms that, algorithmic collusion can be a creditable threat.
  In application,
  we hope, the models of the algorithm design as well as the experiment environment illustrated in this work,
  can be an incubator or a test bed for researchers and policymakers
  to handle the emerging algorithmic collusion.
\end{abstract}

 \begin{flushleft}
   ~~~~~~~~ Cournot duopoly model \\~~~~~~~~  experimental economics \\~~~~~~~~  game theory \\~~~~~~~~ collusion algorithm design \\ ~~~~~~~~ iterated prisoner's dilemma \\ 
   ~~~~~~~~  antitrust \\~~~~~~~~  mechanism design
 \end{flushleft}

\newpage
\section*{Signature of this work}

\begin{enumerate}
  \item Algorithmic collusion is an emergence concept about antitrust in current AI age. \cite{OECD2017May}  \cite{Varian2018}
  \item There is rare laboratory experiment on algorithm-human collusion,
        although human-human collusion experiments has 50 years' history. \cite{Engel2015Tacit}
  \item Cournot duopoly market, which is a basic model for antitrust study, is employed.
  \item A simple linear algorithm is developed to enforce human to collude in the market. (See Equation \ref{eq:keq})
  \item Algorithmic collusion is, to the best our knowledge, firstly observed in our experiment.
  \item As the theoretical expectation, our algorithm can extort its human rival,
        at the same time, can facilitate the collusion in significant (See Table \ref{tab:AHvsHH}). Algorithmic collusion is driven by the human rival unilaterally optimizing his/her own payoff, so is inevitable.
  \item Provide an exemplificative experimental framework to
        understand algorithmic collusion, which can be an incubator for antitrust regulation design.
\end{enumerate}

\newpage
\tableofcontents

\newpage
\section{Introduction}

\subsection{Argument on the creditability of algorithmic collusion threat}
Algorithmic collusion is an emerging concept in current artificial intelligence
age (see Chapter 4 in \cite{OECD2017May}, and \cite{Varian2018} \cite{Capobianco2018}) which was firstly raised by Ezrachi and Stucke in 2015 \cite{Ezrachi2015Artificial}.
Today markets are substantially
determined by algorithmic traders.
Algorithm-driven market agents, however,
poses a new and formidable challenge to existing antitrust laws \cite{OECD2017May}.
If the practice hasn't yet become fully visible for policymakers and regulators,
the Topkins, Tord and Eturas cases suggests that it soon might.
Scholars and policymakers have begun to focus on these potential algorithm threat,
in which algorithmic collusion is a typical threat \cite{OECD2017May}.
In Topkins-like crimes, the collusion is explicit collusion,
because the seller had otherwise demonstrated
a will to collude with other parties
and then coded the algorithm to carry out the agreement.
These collusion
were not established automatically by algorithms, so
Topkins-like algorithm wasn't a real threat.

The real potential threat is
the algorithm, which
can facilitate collusion tacitly with its business rivals
and they together damage the benefit of the whole market and society.
On the threat, the antitrust communities are now split into two groups.
Some of them argue that
it is a actual and a credible threat \cite{OECD2017May,Ballard2017From,Gal2017AFC,EzrachiVirtual}.
Yet, others argue that the possibility is only theoretical conjecture or
something like scientific fictions
 \cite{Deng2017When,Petit2017SIF,Mehra2017SIF,Ittoo2017Algorithmic}, or it is non-credible threat.
In economics science, the argument can be simplified as a scientific question:
When involving market competitions,
whether algorithmic agents can really facilitate tacit collusion?


\subsection{No experimental evidences to settle the argument}
 Facing such questions
we turn to seek scientific evidences from laboratory economic experiments.
Theoretically, there are many strategies that can be used
to facilitate collusion \cite{Varian2018}.
For more than 50 years, economists have been doing experiments
on the tacit collusion in human-human laboratory markets 
\cite{Engel2015Tacit}\cite{Friedman1963Individual}
\cite{Cox1998Learning}\cite{Deck2000Interactions}.
But in algorithm-human competing markets,
rare collusion confirmed empirically.
%
In order to investigate the collusion,
experimental economics has served as a test-bed
for the market dynamics analysis,
and as a incubator for market policymaking
\cite{Plott1982Industrial}\cite{Plott2009Tacit}\cite{Plott2014Public}.
So, evidences from controlled laboratory experiments are expected. 

To assessing the collusion,
Cournot market model is a benchmark \cite{Vives1989Cournot}\cite{Mehra2016AntitrustRobo}.
Meanwhile, duopoly is the simplest case in oligopoly market.
This model has been extensively studied on the economic behavior of \emph{industrial organization} \cite{Tirole1988The}.
As the first step to study the algorithm-human collusion,
we focus on the Cournot duopoly model.
In this model, no algorithm-human collusion has been found till now in experiment.
%
And, in this paper, we hope to provide an experimental evidence of the tacit collusion.
 %
 We hope to see the evolutionary processes of the tacit collusion behaviors,
 the rising of the price, and the declines of the social warfare.
 With these experiment observations,
 we hope to illustrate how
 an algorithmic collusion could be a creditable
 threat | This is the aim of this investigation.

\subsection{Key literature background}

To this aim, our research question can be specified as | how to design an explicit algorithm,
which can practically enforce a human rival to collude in laboratory experiment.
Our research is inspired by two literature,
one is experimental \cite{Wang2016Extortion} and
another is theoretical \cite{Mcavoy2016Autocratic},
both of which base on the zero-determinant (ZD) strategy theory
 in iterated prisoner's dilemma (IPD) \cite{Pressa2012IPD}.
The ZD strategies are a new class of memory-one,
probabilistic and conditional response strategies
that are able to unilaterally
set the expected payoff of its opponent
in the IPD irrespective of the opponent's strategy (coercive strategies).
ZD strategist simplify its opponent decision making
| from strategy interaction problem to a optimal problem.
According to ZD theory, in an algorithm-human interaction IPD game,
if the algorithm applying ZD strategy,
its rational human rival has to cooperate fully.
The ZD theory is not only a theoretical imagine,
it has been supported in long-run laboratory IPD experiments,
in which the human rivals were enforce to cooperate \cite{Wang2016Extortion}.
In basic microeconomics theory,
a Cournot duopoly market game is equivalent to a IPD game
by extending the IPD strategy space to continuous action space.
In continuous action space game,
the theoretical results \cite{Mcavoy2016Autocratic} has shown the existence of extortion strategy.
So, we can expect to demonstrate an algorithm,
which can extort its human rival to collude
in an algorithmic-human experiment.
%

%

\subsection{Our method and result}
%
We investigate the  algorithmic collusion threat by two steps
| First, we proposed an algorithm, which can enforce its human rival to collude; 
Second, we verified the efficiency of the algorithm in laboratory algorithm-human experiments.
In the empirical data,
we see the dynamic (evolutionary) process of the collusion and the constantly decline of the social welfare.
With the data, we confirm that, the algorithmic collusion is creditable threat.
Our contributions are mainly three folds.
(1) We approach an extortion algorithm in duopoly competition,
and illustrate how to design such algorithm explicitly.
(2) We analyzed the efficiency of the extortion algorithm in experiments,
and showed the algorithm can cause great welfare loss in duopoly competition.
(3) More import, our experiment methods can be applied
by antitrust regulators to investigate such algorithmic collusion.
A legal (or ethical) dilemma caused by the algorithmic collusion is discussed at the last.

\section{Model and Experiment Design}
\subsection{Cournot duopoly model and collusion}
Cournot duopoly model is a basic economic model for business competitions \cite{Tirole1988The}.
In this model there are only two agents (firms)
to make a same product, and the competition is on quantity.
The competition is to make decisions how many production
they offer at the same time.
If both firms collude with the other,
they can reduce product supply to a quantity (denoted as $Q_c$) to make their profit biggest.
The Nash equilibrium quantity (denoted as $Q_n$) in this model is a situation
in which firms interact with each other and
choose their best strategy given the strategies that the other have chosen.
If the agents maximum the quantity, the price will be lowest and social welfare will be maximized,
and Walrasian Equilibrium point  (denoted as $Q_w$) will reach.

In Cournot duopoly market, in dynamics view,
the competition is lasting along time.
The strategy and the outcome is keeping changing along time.
In each iteration, each player $i$
chooses a quantity $X_i$ in a finite interval $[L, U]$.
Price $P$ is a decreasing function of the aggregate quantity $z = x + y $, and player $i$'s profit, $S^i$, in that iteration is
 \begin{equation}\label{eq:dan}
    S^i = a + ( P(z) - c ) X_i
 \end{equation}
 including the constant marginal cost $c \geq 0$,
 as well as an exogenous additive constant $a$ that captures benefits
 from other activities net of fixed cost \cite{Friedman2015From}.
 Collusion, as a concept in economics and in legal,
 can be defined with this model \cite{Tirole1988The}\cite{Friedman2015From}.

 In this study case, no loss generality, as an example and the comparability, we apply the exactly same parameters
  setting as the previous human-human
  Cournot duopoly  experiments \cite{Friedman2015From}.
 The parameters are specified as
 \begin{equation}\label{eq:parameter}
   a = c = 10; ~~~~~P(z) = 120 / (x+y); ~~~~~[L, U] = [0.1, 6]
 \end{equation}
 in which
 $L$ ($U$) is the low (up) bound of the available quantity (strategy) of the competitors.
In a duopoly market with these specified parameter, in one round of competition,
if both players choose a quantity of 3 ($Q_n$ = 3),
their profits $S_n$ are 40, respectively,
which implies the price in market is 20 ($P_n$ = 20). At this condition,
neither of the players can benefit by unilaterally deviating this quantity. This state is the Nash equilibrium.
However, if both players choose a quantity of 0.1 ($Q_c$ = 0.1),
they will reach joint profit maximum (JPM) state and get 69 profit respectively.
In JPM state, the price in market is 600 ($P_c$ = 600),
much higher than Nash equilibrium price ($P_n$ = 20), which would harm consumer benefit and social welfare.
So, in economists and market regulators view \cite{OECD2017May}, the JPM state is the state of collusion.

\begin{figure}
\begin{center}
    \includegraphics[angle=0,width=0.6\textwidth]{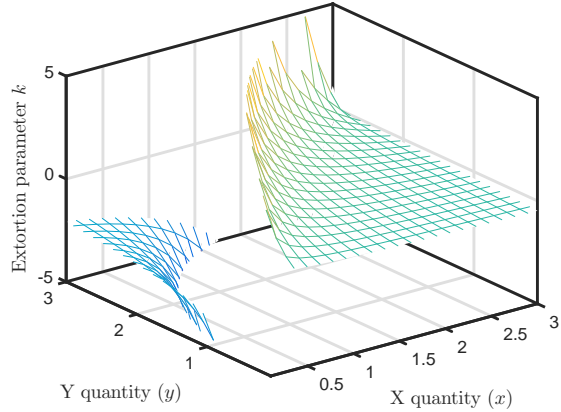}
\end{center}
\caption{
  \textbf{Relationship between the duopoly quantities $x$ and $y$ and extortion parameter $k$.}
  $x$ axis means the quantity of the human rival X.
  $y$ axis means the quantity of the algorithm Y.
  Vertical axis $k$ means normalization ratio between
  the algorithm profit and its human rival profit.
  When $k$ is given, the $y$ is determined by $x$, as shown in Equation (\ref{eq:keq}),
  with the parameters specified in Equation (\ref{eq:parameter}).
  Interval of both $x$ and $y$ are between collusive point and Nash equilibrium point, [0.1, 3].
  It is clear that, when the algorithmic agent determines $k$, for any $x$,
  the algorithm strategy $y$ is determined.
  \label{fig:figa1}
}
\end{figure}
%
\subsection{Linear extortion to collusion algorithm}
To study the algorithmic collusion, we need to design algorithm first.
Considering the equivalent of Cournot duopoly model
and iterated prisoner's dilemma (IPD) game,
our algorithm design is inspired by IPD.
%
%
A class of IPD strategy is described by the outcomes of the previous round,
namely memory one strategy.
%
%
Memory one strategy states that, in a long rounds time
(e.g., day, hour, minute) series $T=\{1,2,3, ..., t-1, t, t+1, ...\}$,
the strategy at $t$ round depends only on the latest previous round ($t-1$).
In a (X, Y) two player game,
at $t$, the player Y can use the information set
at time $t-1$ to decide its own strategy $y$ at $t$ as
\begin{equation}
y_t = g\left(x_{t\!-\!1}, y_{t-1}, S^X_{t\!-\!1}, S^Y_{t-1}\right)
\label{eq:companya}
\end{equation}
We suggest an algorithm, namely Linear Extortion to Collusion Algorithm (LECA).
We set the algorithm Y's normalized payoff $(S^Y - S_n)$ linearly depending on its human rival X's normalized payoff $(S^X -S_n)$ as
\begin{equation}
     S^Y(x_{t\!-\!1},y_{t})-S_n = k\, \left[S^X(x_{t\!-\!1},y_{t})- S_n\right],
    \label{eq:keq}
\end{equation}
in which $k$ is a constant, $y_{t}$ is the quantity of the algorithm Y at $t$, $x_{t\!-\!1}$ is the quantity of the human rival X at $t\!-\!1$,
and $S_n$ is the profile at Nash equilibrium. When $ k > 1$, 
the expected profile of the algorithm Y is to earn more profile than its rival X, 
and Y can be called as extorter.
$k$ is called as the extortion parameter, 
which can be unilaterally determined by the algorithm agent Y. 
For a given $k$, in general, for any given $x_{t\!-\!1}$ (the quantity of the human rival X at $t\!-\!1$), 
mathematically, there could exist a solution for $y_{t}$ (the quantity of the algorithm Y), 
as shown in Figure \ref{fig:figa1}.

So, if there is a stationary strategy solution
($\lim_{t\rightarrow \infty} x_{t} = x_{t\!-\!1} $)
for $x$ , then $y$ is determined.
X can unilaterally optimize its own payoff by solving following equation,
\begin{eqnarray}
  S^X_o &=& \max_{[x_L,\, x_U]} \lim_{t\rightarrow \infty} S^X(x_t,y(x_{t\!-\!1}))
\end{eqnarray}
In general format of the supply-demand function shown in Equation (\ref{eq:dan}),
Y's quantity strategy ($y$) can be numerical solved referring to Equation (\ref{eq:keq}).
As a result, the stationary solution of X's quantity strategy, $x_c$, can be obtained.
The procedure for the algorithm for stationary solution is illustrated in Supplementary Information {}J.
Till now, we have the stationary strategy solution.


However,  we need to think of the game as being played over a number of time
periods (denoted as $N$-rounds), making it dynamic.
Suppose X taking a $N$ step period loop sequence, e.g.,
 $$..., x_{m}, x_{m+1}, x_{m+2}, ..., x_{m+N-1}, x_{m}, ...$$
the related strategy sequence of Y should be
 $$..., y_{m}, y_{m+1}, y_{m+2}, ..., y_{m+N-1}, y_{m}, ...$$
 As the strategy sequence of Y is determined on the X sequence,
 so the outcome is defined. For the human rival, the average payoff of the $N$-step period loop (denoted as $\bar{S_x^N}$)
 can be expressed as
\begin{eqnarray}\label{eq:N}
  \bar{S^X_N } &=& \frac{1}{N} \sum_{t=1}^N S^X_t  = \frac{1}{N} S^X(x_1, y(x_N))  + \frac{1}{N} \sum_{t=2}^N S^X(x_t, y(x_{t\!-\!1})),
\end{eqnarray}
in which $y_t(x_{t\!-\!1})$ is the Y's strategy $y$ (at $t$) in response to X's strategy $x$
(at $t-1$).

  We need to constrain the $k$ value to make sure  $\bar{S_x^N}$ smaller than the payoff $S^X$ when X stop at $x_c$, the stationary solution of X's quantity strategy. So, mathematical question turns to evaluate
whether the maximum of $\bar{S^X_N}$ in $X^N$ space is larger than the stationary solution $S^X_o$. This requires that, for any $N$, the solution for the equation,
\begin{equation}\label{eq:crit}
    \bar{S^X_N } \geq S^X_c
\end{equation}
is an empty set.
That is to require,
in the space $(x_i | i \in \{1,2,...,N\} \, \text{and} \, x_i \, \in [L, U])$,
there is no vector $x \, = (\, {x_1, x_2, ..., x_N})$ to fulfill
the Equation (\ref{eq:crit}), except $x_1 = x_2 = ... = x_N = x^o$.
Once this requirement is fulfilled,
with general hypothesis of rationality, the collusion is inevitable.
In another words, this is the sufficient condition for algorithmic collusion.
In current condition, we do not know the general solution for arbitrary $N$.
In this study, we consider only the first two order, that is $N = 2$ condition.
The procedure is illustrated in Supplementary Information {}K.



\subsection{An application example}
In this study case, by solving Equation (\ref{eq:keq}) with the specification of Equation (\ref{eq:parameter}),  
Y's strategy at $t$ ($y_t$) response to the X's strategy at $t\!-\!1$ $(x_{t\!-\!1})$ is determined, as
\begin{equation}\label{eq:yt}
 y_{t} = \frac{1}{2} \left( k\, x_{t\!-\!1} + 3\, k - x_{t\!-\!1} + 9    \pm {Z} \right)
\end{equation}
 in which $Z_{t-1} = \sqrt{(k + 1)^2\, x_{t\!-\!1}^2 + 9 (k+3)^2  + 6\, x_{t\!-\!1} (k-5)(k+1) }$.
 The optimal strategy (production quantity) of X,  by Equation (\ref{eq:yt}) 
 and Equation (\ref{eq:dan}) with the specification of Equation (\ref{eq:parameter}), can be expressed as
\begin{eqnarray}
       x_c  &=& \arg \max_{x\, \in \, [0.1,\, 6]}  \left(x\, \left(\frac{240}{ {3\, k}  +  {x}  +  {k\, x}  -  {Z}  +  {9} } - 10\right) + 10 \right)
\end{eqnarray}
in which $Z = \sqrt{(k + 1)^2\, x^2  + 6\, (k-5)(k+1) x + 9 (k+3)^2}$.
It is visible that, when $x \in [0.1, 6]$, the optimized unique solution of the human rival X is $x_c =0.1$;
meanwhile, the price is $\left(\frac{240}{ {3\, k}  +  {x_c}  +  {k\, x_c}  -  {Z}  +  {9} } - 10\right)$,
and production quantity of Y is
$y_c = ({3\, k}  - x_c  +  {k\, x_c}  -  {Z}  +  {9})/2$.
The explicit formula and figures are shown in
Supplementary Information {}A.
In $k=1$ condition, the solution go back to quantity pair
 $(x, y)=(0.1000, 0.1000)$, i.e. the JMP state,
 in which profile pair $(S^X, S^Y)=(69.000, 69.000)$, the traditional collusion.

This algorithm can go further. For example,
setting $k=1.2$, the solution of production pair will be $(x, y)=(0.1000, 0.1093)$,
in which profile pair $(S^X, S^Y)=(66.321, 71.585)$, an extortion collusion.
In our experiment, we set $k=1.296$ and
the theoretical extortion collusion pairs are
$(x, y)=(0.1000, 0.1135)$ and $(S^X, S^Y)=(65.203, 72.662)$.
So, referring to Nash equilibrium, the surplus value pair, for human and algorithm respectively, is (25.203, 32.666). This is unfair but the most profitable for the human rival. So the collusion will still be established by the algorithm | This is the theoretical  prediction, which will be test in our experiment.

\begin{figure}[H]
    \centering
    \begin{subfloat}
        \centering
        \includegraphics[width=0.45\textwidth]{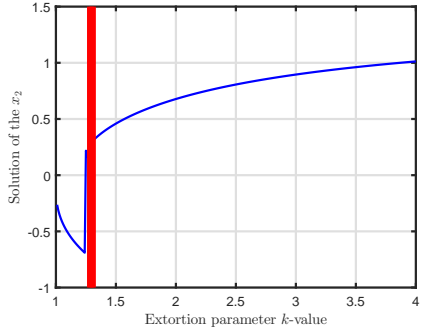}
    \end{subfloat}
    \caption{
    \textbf{First order dynamics solution}
    An illustration of a potential [$x_1, x_2$] jumping
    to deviate the $[x_1,x_2] =[0.1,0.1]$ condition.
    When $x_1 =0.1$, the curve illustrates how
    $x_2$-value depends on the extortion parameter $k$-value, with $x_2 \in [0.1, 6]$. If $k$ = 3, $\bar{S_X^2}[0.1,0.9] > \bar{S_X^2}[0.1,0.1]$, so the jumping is more profitable, and the stationary $x=0.1$ is not the unique solution.
    The red line is the up bound of $k$ (1.296). when $k$ larger than the up bound, the solution of $x_2$ is shown in the blue line.
    }
    \label{fig:x2k}
\end{figure}

As mentioned above, to consider the dynamics processes is necessary. Referring to Equation (\ref{eq:N}),
in the first order dynamics $N$=2 consideration,
\begin{eqnarray}
  \bar{S_x^2} &=& - 5\, x_{1} + \frac{120\, x_{1}}{3\, k + 2\, x_{1} - x_{2} + k\, x_{2} - Z_2 + 9} + 5\\
\nonumber    & &           - 5\, x_{2} + \frac{120\, x_{2}}{3\, k - x_{1} + 2\, x_{2} + k\, x_{1} - Z_1 + 9} + 5
\end{eqnarray}
in which $Z_i = \sqrt{(k + 1)^2\, x_i^2  + 6\, (k-5)(k+1) x_i + 9 (k+3)^2}$ and $i \in \{1,2\}$.
So, for any given $k$, if there exists $(x_1, x_2) \in [0.1, 6]^2$
to satisfy Equation (\ref{eq:crit}), such $k$ is not valid.
Analysis result shows that, the valid extorter parameter
$k$ value is
\begin{equation}\label{eq:kinterval}
  1 < k < 1.296
\end{equation}
The human rival could benefit from deviation, if $k> 1.296$.
As a numerical example, assume $k = 3.0$,
a strategy which can be used to resist our algorithm,
the human rival may take the jumping strategy
\{0.1, 0.9, 0.1, 0.9...\} (with profile 59.9146) instead of the strategy
\{0.1, 0.1, 0.1, 0.1...\} (with profile 54.3383). So, $k=3.0$ is not fit.
The deviation strategy for $N$ = 2 condition can be calculated analytically
(explicate results are shown in Supplementary Information {}B),
and figural results are shown in Figure \ref{fig:x2k}.

%
%
%


\subsection{Experiment System}

\begin{figure}[H]
	\begin{center}
    		\includegraphics[angle=0,width=0.45\textwidth]{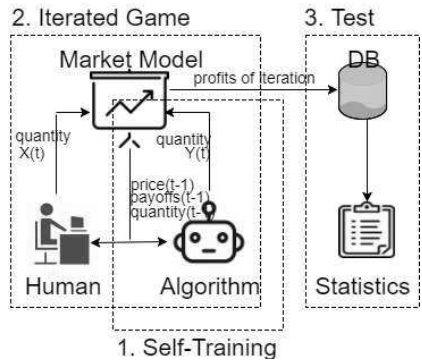}
	\end{center}
	\caption{
	\textbf{Experiment Framework:} This framework consists three phases: \textbf{Phase1. Self-Training:} Giving algorithm the market model, algorithm calculate some parameter by testing case. \textbf{Phase2. Iterated Game:} Recruit human subjects to compete with algorithm. The processes have already described above. \textbf{Phase3. Test: }The information about iterated game will been stored in database. With the data, researchers can evaluate the algorithmic collusion. \label{fig:fig1}
	}
\end{figure}

For evaluating the performance of LECA,
 we designed an experiment system. The whole system consists three phases shown in Figure \ref{fig:fig1}. \textbf{Phase-1  Self-Training:} The Self-Training is to calculate some parameters in LECA (already calculated in \textbf{First order dynamics solution} section).  \textbf{Phase-2 Iterated Game:} The Iterated Game is providing a procedure which every human subjects faced with  the LECA implemented by a computer for many iterations in the fixed experiments settings. In each iteration, LECA and human subjects can select a quantity of products they want to produce, then the Market Model calculates the human subject's and LECA's profits and feedback to human subject and LECA. At the same time, the server save the information into database. After checking their strategy and profit, LECA and its human subject will decide their strategy (quantity) of the next iteration. \textbf{Phase-3 Test:} Test is analyzing the behavior of human subjects to evaluate the efficiency of algorithm (see Result section).

 To make experiment simple, LECA algorithm is stored in referee (server), the quantity LECA produce is calculated by referee. In our experiment, the interval of quantity is [0.1, 3] for the algorithm and [0.1, 6] for its human rival, and the extortion parameter is $ k = 1.296$. More details of the technology detail, as well as the difference of the experimental protocols comparing with \cite{Friedman2015From}, are shown in Supplementary Information {}C.

\begin{figure}[H]
\begin{center}
    \includegraphics[angle=0,width=0.9\textwidth]{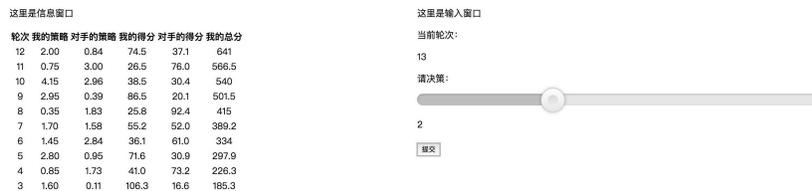}
\end{center}
\caption{
  \textbf{Experiment UI} An illustration of the asset market task. The left panel indicates the history information of 10 iterations. After submitting the quantity (right panel), the server will calculate the quantity and the profits, and the result will shown in the first line. Those result keep for 10 iterations. The right panel is decision making panel, a slider which can be dragged represents the quantity that human subject want to produce. After decision was made, human subject can click the button to submit the decision.\label{fig:fig2}
}
\end{figure}

In an iterated game, a human subject will face the user interface (UI) shown in Figure \ref{fig:fig2}.
There are two panels.
The left panel shows the information
about the last ten iterations of the game,
including 6 items: the number of the round  (iteration) in the experiment,
human subject's quantity, LECA's quantity,
human subject's profit, LECA's profit
and human subject's total profit, respectively.
The right panel is the decision making panel.
In each round, human subject can make decision by submitting the quantity.
When submission done,  the server reply information of the 6 items,
and a new round starts again.

\subsection{Experiment  Procedure}
The data we used here come from our laboratory experiments which were conducted at Experimental Social Science Laboratory of Zhejiang University, on 25 December 2017 and 16 to 18 January 2018. A total of 40 undergraduate and graduate students from various disciplines were recruited to participate in the experiment with each of them only participating once.
In total, we collected 24000 observations of individual decision making, consisting of choices of human subjects and LECA algorithm implemented by computer programs. Each experiment lasted for 1 hour. During the experiment, the player earned scores according to Experiment Environment section and their choices. After the experiment, the sum of scores were converted to cash according to an exchange rate and paid to the subjects, the converted function is:
\begin{center}
$1.2 \times (\text{TotalProfit} / 600 - 30) + 5$ ~~~~~~~~~~(unit in Yuan RMB)
\end{center}
Here,  $\text{TotalProfit}/600$ Yuan RMB is average profit that human subjects earn,
30 Yuan RMB is baseline, 1.2 is a rate, and 5 Yuan RMB is show-up fee.
---- In the 600 rounds If the two players fall into Nash equilibrium,
both of our algorithm and the human rival gain 24000 points; If
the human stays in the smallest value (low bound) of
the state space ($x = 0.1$) , the human will earn 39660 points total.
%
The average earning is about 40 Yuan RMB.\par
Before the formal experiment, every human subjects will be allocated a computer.
They were then assigned an instruction manual and a pen,
and they played the game in an small isolated room with a computer.
No oral instructions were given,
except that the organizer told the subjects
(1) not to refresh the web page to avoid potential technologic problem (the experiment user interface is a web page), and (2) the type error in the one page printed Experiment Instruction (The Experiment Introduction for the student subjects is attached in Supplementary Information {}D).
They made decisions by dragging the slider on the screen and clicking submitting button. The software for the experiment was designed by the authors. No communication was allowed. The human subjects were told that they would play a game with a fixed computer program for 600 rounds. The algorithm code is attached in Supplementary Information {}E.

\section{Result}

Algorithm-human collusion is observed in our experiment.
We see the supplement constantly decreases,
while the human rivals accept the extortion.
We see the degree of collusion constantly increases,
while the social warfare constantly decreases.
Comparing with existing human-human experiment,
the algorithm-human collusion can be more harmful to society.
All these observations indicate that,
not theoretical conjectures or scientific fictions,
the algorithmic collusion is a creditable threat.

\subsection{Human rivals collude with the algorithm}
\begin{figure}[H]
\begin{center}
    \includegraphics[angle=0,width=0.5\textwidth]{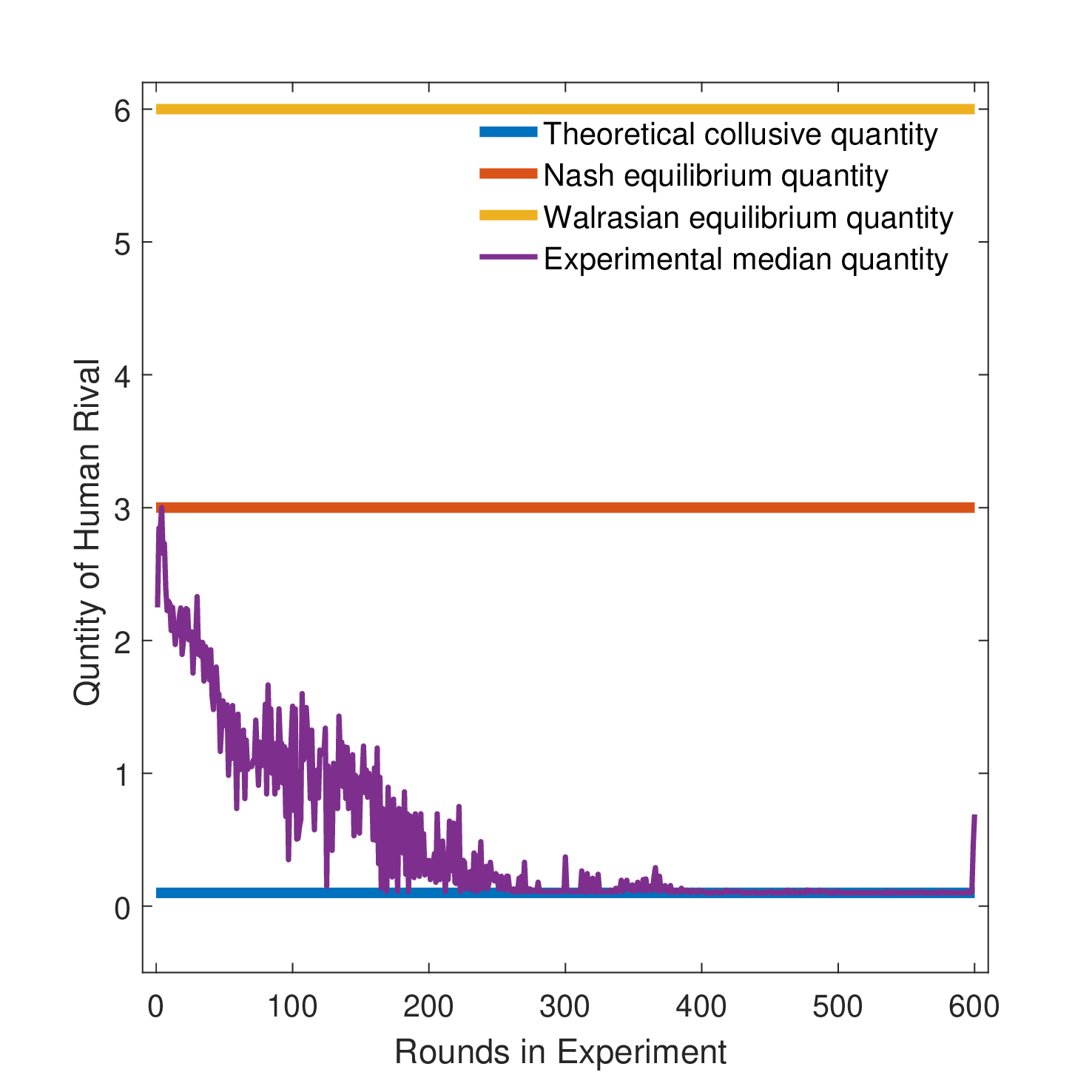} 
\end{center}
\caption{
  \textbf{Median quantity of human subjects:} The purple line shows the median quantity of 40 samples. Along time, the quantity decline to the blue line (theoretical collusive line). There are three theoretical lines, represent the quantity that human subjects decide to collude, to keep at Nash equilibrium and to keep at the Walrasian equilibrium, respectively. 
  Here, the Walrasian equilibrium means both players fully competing, which both can not get any profit from the competition.
  \label{fig:fig3}
}
\end{figure}

Collusion, in a Cournot duopoly market, can be identified by the constantly declining of the total supplement (quantity).
Figure \ref{fig:fig3}, reporting the median of the quantity of the 40 human rivals of our algorithm, shows clearly that in first 300 iterations of Cournot game, the human subjects reduce quantity to reach almost fully collusive level (0.1), then keep their quantity at the collusive level.
This result indicates that after about 300 iterations of learning,
the human subjects realized they will get biggest profits when they select a quantity of 0.1 (collusive level).
For more details on the evolution of the individual behaviors, see Supplementary Information {}M.

To make things more clear, Table \ref{tab:tab1} shows the average,
standard deviation and median number of quantity and profit 40
human subjects' performance in iteration 1 - 600,
iteration 1 - 300 and iteration 301 - 600.
Table \ref{tab:tab1} shows that, at the beginning (iteration 1 to iteration 300), the average and median of human subjects' quantity is much more larger than the later iterations (iteration 301 to 600). That means, with the increase of round, human subjects reduce quantity of product to cooperate with LECA. In the mean time, the profits human subjects earn raise apparently.

\begin{table}[ht]
\begin{center}
  \caption{Quantity and profit of human subject}\label{tab:tab1}
  \scriptsize
\begin{tabular}{r|rrr|rrr|r}
    \hline
    iteration&&quantity&&&profit  \\
    \cline{2-7}
    &average&stdev&median&average&stdev&median\\
    \hline
    1-600&1.182&1.216&0.664&57.168&9.453&61.768 \\
    1-300&1.442&1.265&1.044&55.034&9.575&58.771 \\
    301-600&0.923&1.251&0.265&59.302&9.923&64.478 \\
    \hline
\end{tabular}
\begin{tablenotes}
There were 40 samples (student subjects) participated the experiment.
Each subject has competed with the algorithm for 600 iterations.
The procedure to reach the results shown in this table as following:
At first, calculate each sample's average quantity and average profit of 600 iterations, first 300 iterations and last 300 iterations. Then, calculate the average, standard deviation (stdev) and median of those samples.  Take median quantity for example, $0.664$ is the median number of 40 samples' average quantity in iteration $1$ to iteration $600$. $1.044$ is the median number of 40 samples' average quantity in iteration $1$ to iteration $300$. $0.265$ is the median number of 40 samples' average quantity in iteration $301$ to iteration $600$.
\end{tablenotes}
\end{center}
\end{table}
\par

\subsection{Algorithm extorts its human rival and earn higher profile}

Now we show LECA should get more profit than its human rival in the algorithm-human collusion.
Table \ref{tab:tab2} show the statistical comparison results about the profits of LECA, human subjects and Nash Equilibrium. The statistical analysis methods are  interpreted in Supplementary Information {}L.
\begin{table}[ht]
\begin{center}
  \caption{Statistical significant ($p$-value) of the profiles comparisons$^*$}\label{tab:tab2}
  \scriptsize
\begin{tabular}{|r|rr|rr|rr|}
    \hline
    Hypothesis &Algorithm&$>$ ~~~Human~~~~&Algorithm&$>$ ~~~Nash~~~~~~&Human&$>$~~~~~Nash~~~~\\
    \cline{2-7}
    Iteration&rs&tt&rs&tt&rs&tt \\
    \hline
    1-600&0.0177 &9.77$\times 10^{-07}$&6.97$\times 10^{-10}$&1.34$\times 10^{-12}$&4.76$\times 10^{-11}$&4.39$\times 10^{-14}$ \\
    1-300&0.142 &7.24$\times 10^{-05}$&6.97$\times 10^{-10}$&1.25$\times 10^{-10}$&6.97$\times 10^{-10}$&3.11$\times 10^{-12}$ \\
    301-600&1.71$\times 10^{-04}$&5.77$\times 10^{-08}$&4.45$\times 10^{-10}$&1.18$\times 10^{-13}$&5.04$\times 10^{-12}$&5.31$\times 10^{-15}$ \\
    \hline
\end{tabular}
\begin{tablenotes}
* Algorithm here is our Linear Extortion to Collusion Algorithm (LECA). Human here is the human rivals of LECA in our experiment.
Nash indicates the theoretical profile (40) when both duopoly employ Nash equilibrium at per iteration.
$rs$ represents rank-sum: returns the $p$-value of a two-sided Wilcoxon rank sum test.
$tt$ represents $t$-$test$: returns the $p$-value of testing the null hypothesis that the pairwise difference between data vectors has a equal mean. Procedure to report the $p$ values here is:
At first, calculate the profile of Algorithm and Human of all 40 samples, respectively. Then, calculate statistical results of the profile comparison with the null hypothesis that (1) Algorithm $=$ Human, (2) Algorithm $=$ Nash and (3) Human $=$ Nash, respectively.
\end{tablenotes}
\end{center}
\end{table}
Theoretically, we can easily find that LECA's profit is more than its rival when rival's quantity stay unchanged. In this experiment, the result is significant even though human subject's quantity changes.
From Table \ref{tab:tab2}, we can come to a conclusion that the profit they get is
$$\text{Profit(Algorithm)} > \text{Profit(human)} > \text{Profit(Nash equilibrium)},$$
because $p$-value of the ranksum and t-test is very small and closes near to 0. Interestingly, in the table, iteration 1 to 300, the $p$-value of ranksum and t-test is much more bigger than iteration 1 to 600 and iteration 300 to 600. In iteration 1 to 300, there still exist human subjects who want to compete with LECA to gain more profits. However, after finding that is impossible, more and more human subjects begin to cooperate with LECA even though they are extorted by LECA. This result confirms that Human subjects cooperate with LECA.

\subsection{Degree of collusion}
\begin{figure}[H]
\begin{center}
    \includegraphics[angle=0,width=0.5\textwidth]{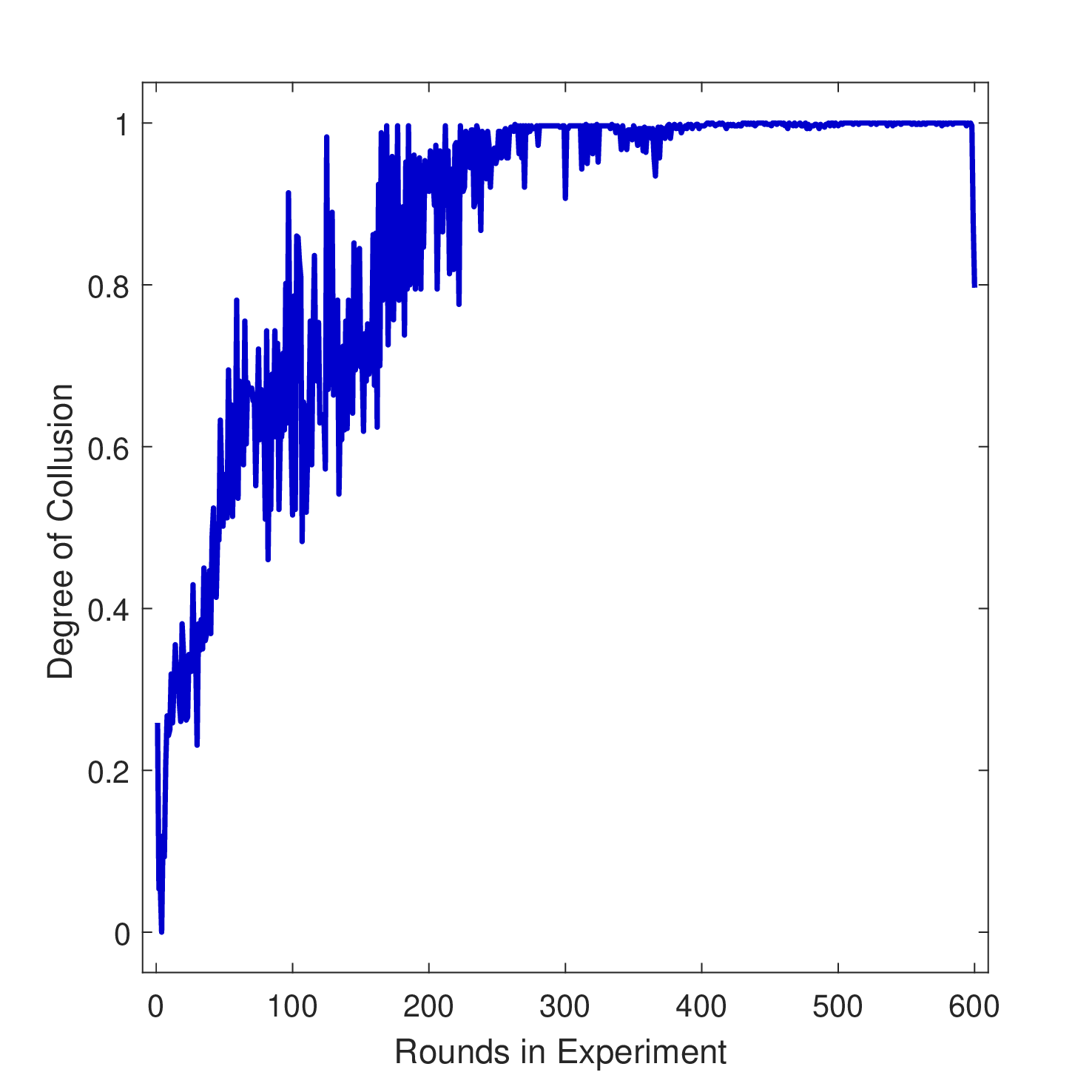} 
\end{center}
\caption{
  \textbf{Degree of Collude: } Degree of Collude can be described as $\frac{N_q - x}{N_q - C_q}$, where $N_q$ means the quantity when both oligopolists reach Nash equilibrium level, $x$ means the quantity that human subject produce, $C_q$ means the quantity when both oligopolists reach collude level. This figure is the median degree of collude in each round for all 40 samples.
In our experiment, the theoretical maximum degree of collusion value is $(6-0.21)/(6-0.2)$ = 0.998, which is slightly lower than 1, because of algorithm extortion motivation, see Equation (\ref{eq:keq}).
  \label{figa:fig3}
}
\end{figure}

Degree of collusion is a standard measurement
for the anti-competition behavior.
This measurement can be expressed as a percentage of
the distance between the Nash and the Pareto (collusion) outcomes \cite{Engel2007HOW}.
In our Algorithm-Human duopoly market,
the evolution of the degree of collusion is shown in Figure (\ref{figa:fig3}).
It is clear that, in our Algorithm-Human duopoly market,
the degree of collusion rises to nearly 100\% in 300-400 rounds.

\subsection{Deadweight loss}
Deadweight loss is a loss of economic efficiency that
can occur when equilibrium for a good or a service is not achieved.
In our study case, the deadweight loss (or social welfare lose), at time $t$, can be formulated as
$\int_b^a (D(x)-S(x))\,dx$, in which $a = x_t + y_t$
(the sum of the quantities of the algorithm $y_t$ and
of its human rival $x_t$), $b = 9$
(sum of the up bound of the quantities of the algorithm (3) and of it human rival (6)),
$S = 10$ (the constant marginal cost $c$ in Equation (\ref{eq:parameter}))
and
$D(x) = 120/x$ (the price function, see Equation (\ref{eq:parameter})).
In our study case, the theoretical expectation of the deadweight loss is
$\int_{0.21}^9 (120/x -10)\,dx$ = 363.0447  (367.5665278 For more details, see Supplementary Information {}F).
The red line in Figure (\ref{fig:fig4}) illustrates the deadweight loss along time.
\par
\begin{figure}[H]
\begin{center}
    \includegraphics[angle=0,width=0.5\textwidth]{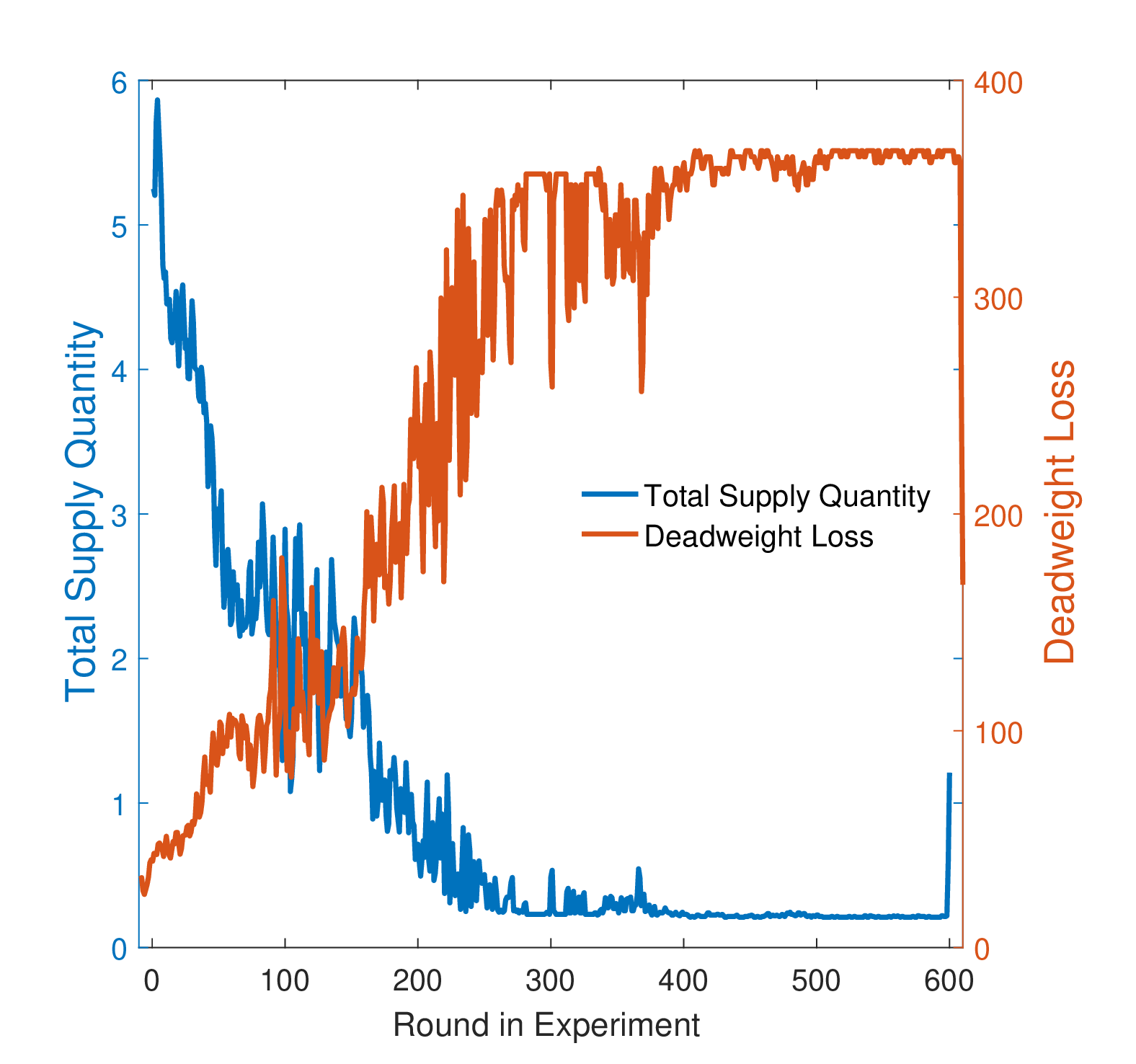} 
\end{center}
\caption{
  \textbf{ Evolution of the duopoly output quantity (in blue) and the deadweight loss (in red)}. Median quantity of both oligopolists and deadweight loss in the 600 iterations. Horizon axis indicates the iteration number; Left vertical axis indicates the sum of the median number of human and LECA's total quantity in the iteration;
  Right vertical axis indicates the deadweight-loss in the iteration.
  Median quantity decline in the first 300 iterations from 5 to 0.2; As result, the deadweight loss increases from 30 to 360 in the first 300 iteration.\label{fig:fig4}
}
\end{figure}
\par

From Figure \ref{fig:fig4}, the quantity of both oligarchies drop down quickly, so, in the Cournot duopoly market, the supplement drop quickly, which leads the increasing of deadweight loss, and is a threat to social welfare.

\subsection{Algorithm-human vs. human-human collusion}
As we apply the exactly same parameters as
the previous human-human Cournot duopoly experiments \cite{Friedman2015From},
it is meaningful to compare our results with theirs.
The median quantities, prices, and profits of human subject in human-human competition is shown in the right panel of Table \ref{tab:AHvsHH} (data comes from the Table 2,
duopoly condition, Page 193 in \cite{Friedman2015From}).
As the results from our experiment, the median quantities, prices, and profits of the algorithm, as well as its human rival, are shown in the left panel of Table \ref{tab:AHvsHH}.

\begin{table}[H]
\begin{center}
  \caption{Comparisons by markets, players, quantities, prices, and profits (median)}\label{tab:AHvsHH}
  \scriptsize
\begin{tabular}{|rrrrrr|rrrr|}
    \hline
             &Algorithm&      &Human &        &      &          &Human& & \\
    \hline
    iteration&Quantity&Profit &Quantity&Profit&Price & iteration&Quantity&Profit& Price\\
    \hline
    1-50&		2.14&	49.79&		2.01&	47.28&	28.57&		1-50&	4.54&		23.74&	13.98\\
    1-200&		1.22&	62.51&		1.10&	57.54&	53.21&		1-400&	3.17&		35.45&	18.43\\
    201-400&		0.14&	67.56&		0.12&	61.16&	436.36&		401-800&	0.57&		63.11&	90.01\\
    401-600&		$^*$\textcolor[rgb]{1.00,0.00,0.00}{0.11}&	$^*$\textcolor[rgb]{1.00,0.00,0.00}{70.30}&		$^*$\textcolor[rgb]{1.00,0.00,0.00}{0.10}&	$^*$\textcolor[rgb]{1.00,0.00,0.00}{64.81}&	558.14&		801-1200&	$^*$0.28&		$^*$68.53&	107.36\\
    551-600&		0.11&	70.30&		0.10&	64.81&	558.14&		1151-1200&	0.40&		68.51&	91.30\\
    \hline
    Theoretical& Algorithm&      &Human &        &      &          &    & & \\
    Expectation$^*$&0.1093 & 72.662& 0.10 &65.203        &   573.34    &        &0.1 & 69 &600 \\
    \hline
\end{tabular}
\begin{tablenotes}
$^*$ In our experiment, the quantities are displayed in two decimal places in the user interface (see Figure \ref{fig:fig2}).
 And then the profiles are calculated referring the displayed quantities.
 So, in the experiment data, the expectations for the algorithm the quantity equals to 0.11 and the profile equals to 71.75;
 Meanwhile the expectations for the human rival, the quantity equals to 0.1 and the profile equals to 66.14.
 The experiment data is provided in the publisher web site,
 and its interpretation is provided in Supplementary Information {}G.
\end{tablenotes}
\end{center}
\end{table}
In Table \ref{tab:AHvsHH}, a significant point is that,
in the later phase in our experiment (400-600 rounds),
the collusion form.
At this condition, the algorithm gain higher profile (70.30)
than both of the theoretical collusion profile (69.00)
and the experimental human-human collusion profile (68.53).
Meanwhile, its human rival gain less (64.80).
In our algorithm-human experiment,
the time to establish collusion (about 400 rounds)
is less than human-human collusion (about 800 rounds) experiment.
By the comparison, two potential results could be:
\begin{enumerate}
  \item Algorithm can facilitate the collusion more quickly.
  \item There exists incentive for a firm to implicate such algorithm in market.
\end{enumerate}

More details results of the comparisons,
especially the individual behaviors of human subject in the experiments,
will be shown in Supplementary Information {}H.

%
%



\section{Discussion}
The question |
the risk that algorithms \textbf{may} work as a facilitating factor for collusion | is deeply concerned \cite{OECD2017May}.
In this paper, we have proposed an algorithm which 
automatically extort its rational rival to collude, meanwhile gets higher profile in the collusion. 
%
%
In laboratory experiments of the algorithm-human Cournot duopoly competitions,
the algorithmic collusion is demonstrated. 
We see the supplement constantly decreases,
the degree of collusion constantly increases,
while the social warfare constantly decreases.
All these observations indicate that,
not theoretical conjectures or scientific fictions,
the algorithmic collusion is a creditable threat.
To the best of our knowledge,
this is the first empirical observation to confirm that
algorithms \textbf{can} work as a facilitating factor for collusion.

\subsection{On algorithm for algorithm-human collusion}
\textbf{How to design an algorithm, which can facilitate collusion by extortion,
 is demonstrated explicitly.}
%
The proposed algorithm (shown in Equation \ref{eq:keq})
is design to enforce its human rival to unilaterally optimizing the payoff,
so the algorithmic collusion is a result of rationality 
and is \emph{inevitable}. 
In this view, the algorithm design likes the mechanism design, 
by which the collusion is determined by the algorithm. 
%
%
The algorithms, which can enforce its rational rival to collude, could be abundant,
from imitation \cite{Friedman2015From} to generously persuade
(e.g., \cite{Pressa2012IPD}\cite{Wang2016Extortion}). 
We have only used memory one strategy, as shown in Equation \ref{eq:companya}. 
Deeper memory strategy is naturally an opinion for further algorithm design.
Considering the equivalent of Cournot duopoly model and IPD game,
the algorithms in IPD game competing \cite{Axelrod1984Evolution}\cite{Pressa2012IPD}\cite{Mcavoy2016Autocratic}
can be adapted to study the algorithmic collusion further.


  The limitation of our algorithm is obvious.
  In our study case, the extortion parameter $k$ is limited (see Equation \ref{eq:kinterval})
  because the human rival could exists dynamic process
  to deviate the collision stationary equilibrium to gain more benefit.
  The theorem of the autocratic strategies \cite{Mcavoy2016Autocratic} provides
  only static solution rather than dynamic,
  but there could exist switching strategy  (dynamic behavior) to against the algorithm.
  In fact we have noticed that, when in higher order dynamic process, e,g,
  in Equation \ref{eq:crit} when $N$ = 3, there exists dynamic solution
  for human rival to deviate the collusion, even at $k$=1 condition. 
  For more details, see Supplementary Information {}I. 
  Equation \ref{eq:crit} provides only the necessary condition, 
  and the sufficient condition is not known.  


    The competition between oligopoly and governor seems inevitable too |
    The oligopoly owner will face how to develop a
    \emph{harder to detect, more effective,
    more stable and persistent} \cite{Ballard2017From} algorithm for business success.
    Meanwhile the regulation (governor) will face
    how to detect such antitrust algorithm to protect human wellbeing.

\subsection{On experiment of algorithm-human collusion}
\textbf{The experimental evidence of algorithm facilitating collusion between human-algorithm is firstly reported.}
Collusion in oligopoly game is standard part in the application of game theory, especially in industry organization theory \cite{Tirole1988The}.
%
Cournot duopoly market, as the dual of Bertrand duopoly market, is the most basic model for study collusion.
Most related experiments are summarized in the survey \cite{Potters2013OLIGOPOLY} and the recent long-run experiment \cite{Friedman2015From}.
The collusion was not observed in short-run experiments (e.g., \cite{Cox1998Learning}), but can be observed in long-run \cite{Friedman2015From}. On this point, we set our experiment times to be long run 600 rounds repeated.
Most previous experiment aimed at human-human interaction, ours is to study algorithm-human collusion.
To our knowledge, the collusion between algorithm-human in Cournot duopoly market is firstly reported here.  \cite{Potters2013OLIGOPOLY} \cite{Friedman2015From}.
%


%

Although we can see the collusion (evolutionary) processes in our experiment,
but the mechanism of the human behaviors is not clear.
Further questions include 
(1) whether the time spent would be shorter or longer in other experimental protocol.
(2) whether or how the collusion
can be established  when a market includes more participates
\cite{Tirole2003The}\cite{Selten1973A}, or (3) when the algorithm and human competitors are mixed,
or (4) when the information environment differs,
or (5) when the frequency of strategy changing is limited. 
It is not surprises that, all old issues need to revisit when algorithms involve \cite{OECD2017May}.
%

\subsection{On dilemma of algorithm-human collusion}
\textbf{The legitimacy of our algorithmic collusion becomes a challenge to current legal system.}
 There are two  approaches, economics and legal,
 to consider whether a supra-competitive price strategies is right or wrong.
 Economists usually distinguish between two forms of collusion,
 explicit and tacit, by the interaction behaviors.
 Explicit collusion is forbidden. But tacit may be allowed,
 %
because, the supra-competitive price strategies may be the normal outcome of rational economic
behaviour of each firm on the market \cite{OECD2017May}.
In our experiments, different from Topkins-like algorithm cases, there is not
any communication or agreement
between the competitors. So the collusion is not explicit,
and is not definitive wrong.

Contrary to the economic approach, which considers collusion a market outcome,
the legal approach focuses on the \textbf{means} used by competitors to achieve such a collusive outcome \cite{OECD2017May}.
The mean of the algorithm approached, shown explicitly in Equation (\ref{eq:keq}) in this paper, appears normal.
Its mean can be explained as the willing to gain $k$-times (in our experiments, $k$=1.296) more profile
than its competitor (the human rival).
Meanwhile, as a rational economic
behaviour, the human is obliged to
collude with the algorithm for the maximizing its own profile.
So, not only its human rival, referring to
current legal consideration, the algorithm is right. As both are right,
as their aggregate behavior, our algorithmic collusion is right?

So, the dilemma is that, the algorithm can enforce the supra-competitive price strategies
and the social human wellbeing is harmed,
but current legal seems to say such algorithm is right.
Recent has seen the establishment in the regulations
on algorithms, requirements includes
transparency and accountability \cite{USACM2017}, as well as ethics
IEEE P7000x \cite{Chatila2017The}.  We wish our work can promote the developments on related fields.

\subsection{Implications of our study}
Our study has a number of implications:
\begin{itemize}
  \item (1) To settle down the argument on whether the threat of algorithmic is creditable.
    Most notably, we have provided  the first evidence in which algorithmic
    collusion automatically arises | by extorting its human rival,
    an algorithm can efficiently facilitate the tacit collusion.
  \item (2) To provide an incubator for antitrust legal research. Pioneered by Charles Plott and his colleagues, in past decades, laboratory experiment has become an incubator and a test-bed for political economics science, e.g., industrial organization \cite{Plott1982Industrial}. As pointed out by Ezrachi and Stucke \cite{Ezrachi2017Two}, the agency would then test what factors (e.g., information, or number of participation, or frequency of strategy changing, or noise) added to (or removed from) the incubator would make tacit collusion likelier and more durable. 
      Our work has provided an explicitly example for algorithm-driven tacit collusion. We hope basing on the algorithm design (e.g., Equation \ref{eq:keq}) and the experiment framework (Figure \ref{fig:fig2}), algorithmic collusion can be investigated more practically.
  \item (3) For algorithmic engineers in business area,
      our study has provided an example to design collusion strategy for business robot.
      However, we hope the engineers, by Figure \ref{fig:fig4},
      can understand the potential harmfulness of algorithmic collusion on social welfare,
      as well as the human wellbeing.
      And then, engineers would consciously obey the AI industry ethical standards,
      e.g., IEEE P7000 series \cite{Chatila2017The}.
\end{itemize}

\end{document}